\documentclass{emulateapj}
\usepackage{graphicx}

\shorttitle{Electron MHD Turbulence}
\shortauthors{Cho \& Lazarian}

\begin{document}

\title{The Anisotropy  of Electron MHD Turbulence}

\author{Jungyeon Cho}
\affil{CITA, Univ. of Toronto, 60 St. George St., Toronto,
              ON M5S 3H8, Canada; cho@cita.utoronto.ca}

\and 

\author{A. Lazarian}
\affil{Astronomy Dept., Univ.~of Wisconsin, Madison, 
       WI53706, USA; lazarian@astro.wisc.edu}

\begin{abstract}
We present numerical studies of 3-dimensional electron magnetohydrodynamic 
(EMHD) turbulence.
We investigate cascade timescale and anisotropy 
of freely decaying strong EMHD turbulence with zero electron skin depth.
Cascade time scales with $k^{-4/3}$.
Our numerical results clearly show scale-dependent anisotropy.
We discuss that the observed anisotropy
is consistent with $k_{\|} \propto k_{\perp}^{1/3}$,
where $k_{\|}$ and $k_{\perp}$ are wave numbers parallel and
perpendicular to (local) mean magnetic field, respectively.

\end{abstract}

                             

\section{\label{sec:intro}Introduction}
Electron magnetohydrodynamics (EMHD)
deals with MHD phenomena occurring at scales smaller than
those of conventional MHD (see Kingsep, Chukbar, \& Yankov 1990).
On scales below the ion inertial length $d_{i}=c/\omega_{pi}$, where
$c$ is the speed of light and $\omega_{pi}$ is the ion plasma frequency,
we can assume that
the ions create only smooth motionless background and fast electron flows
carry all the current, so that
\begin{equation}
  {\bf v}_e 
   = - \frac{ {\bf J} }{ ne } 
            = - \frac{ c }{ 4 \pi n e } \nabla \times {\bf B}. \label{eq1}
\end{equation}
When we ignore the ambipolar diffusion, the Ohm's law becomes
${\bf E} = - { {\bf v}_e \times {\bf B} }/{ c } + \eta {\bf J}$,
where $\eta$ is the plasma resistivity.
Inserting this into Faraday's induction law 
($\partial {\bf B}/\partial t = -c \nabla \times {\bf E}$),
we obtain the EMHD equation
\begin{equation}
   \frac{ \partial {\bf B} }{ \partial t }
  = - \frac{ c }{ 4 \pi n e } \nabla \times \left[
    (\nabla \times {\bf B}) \times{\bf B} \right] + \eta \nabla^{2} {\bf B}.
\end{equation}
Note that, in this paper, we only consider the zero (normalized)
electron inertial length case: $d_e=c/(\omega_{pe}L) \rightarrow 0$,
where $\omega_{pe}$ is the electron plasma frequency and $L$ is the
typical size of the system. 

EMHD turbulence (also known as whistler turbulence)
can be viewed as a generalized
Alfvenic turbulence on scales smaller than
the proton Larmor radius (Quataert \& Gruzinov 1999).
EMHD turbulence plays important roles in neutron stars and
accretion disks.
Goldreich \& Reisenegger (1992) discussed
the properties of EMHD turbulence and argued 
that EMHD turbulence can enhance
ohmic dissipation of magnetic field in isolated neutron stars (see also
Cumming, Arras, \& Zweibel 2004).
Quataert \& Gruzinov (1999) discussed transition from
conventional Alfvenic MHD cascade to EMHD cascade in advection dominated 
accretion flows (ADAF).
Other applications of the EMHD model include
quasi-collisionless magnetic reconnection 
in laboratory and space plasmas (Bulanov, Pegoraro, \& Sakharov 1992;
Biskamp, Schwarz, \& Drake 1995; Avinash et al. 1998) and
plasma opening switches and Z pinches (see Kingsep et al. 1990).

Earlier researchers convincingly showed that
energy spectrum of EMHD turbulence is steeper than 
Kolmogorov's $k^{-5/3}$ spectrum. 
Using two-dimensional (2D) numerical simulations, 
Biskamp, Schwarz, \& Drake (1996) 
found that energy spectrum follows $E(k) \propto k^{-\mu}$ with
$\mu=2.25\pm 0.1$.
They showed that the following Kolmogorov-type
argument can explain this spectrum.
Suppose that the eddy interaction time for eddies of size $l$ is 
the usual eddy turnover time
$t_{cas,l} \sim l/v_l$.
Since ${\bf v}\propto \nabla \times {\bf B}$ (equation (\ref{eq1})),
this becomes $t_{cas,l} \propto l^2/b_l$.
Combining this with the constancy of
spectral energy cascade rate ($b_l^2/t_{cas,l}$=constant),
one obtains 
$E(k)\propto k^{-7/3}$. 
Note that $E(k)$ and $b_l$ are related by $kE(k) \sim b_l^2$.
Biskamp et al. (1999) also obtained a similar result for 3D.
Ng et al. (2003) confirmed the 2D result of 
Biskamp et al. (1996).
Therefore, earlier works strongly support that
\begin{equation}
  E(k) \propto k^{-7/3} \mbox{~~~(as $d_e \rightarrow 0$)}
\end{equation}
for both 2D and 3D.
Recently, Galtier \& Bhattacharjee (2003) studied energy spectrum of
{\it weak} whistler turbulence.


\begin{figure*}[t]
\includegraphics[width=.32\textwidth]{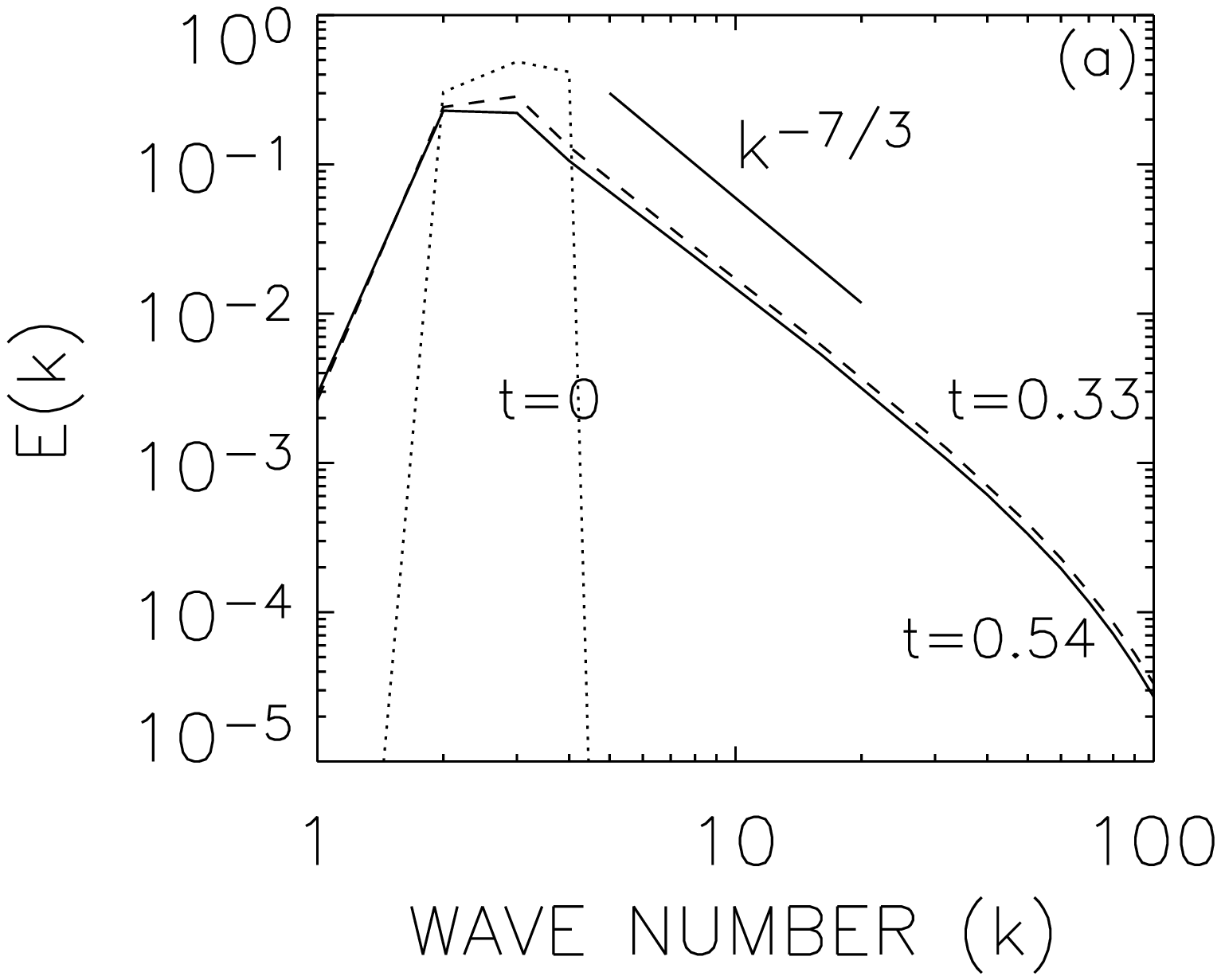}
\includegraphics[width=.32\textwidth]{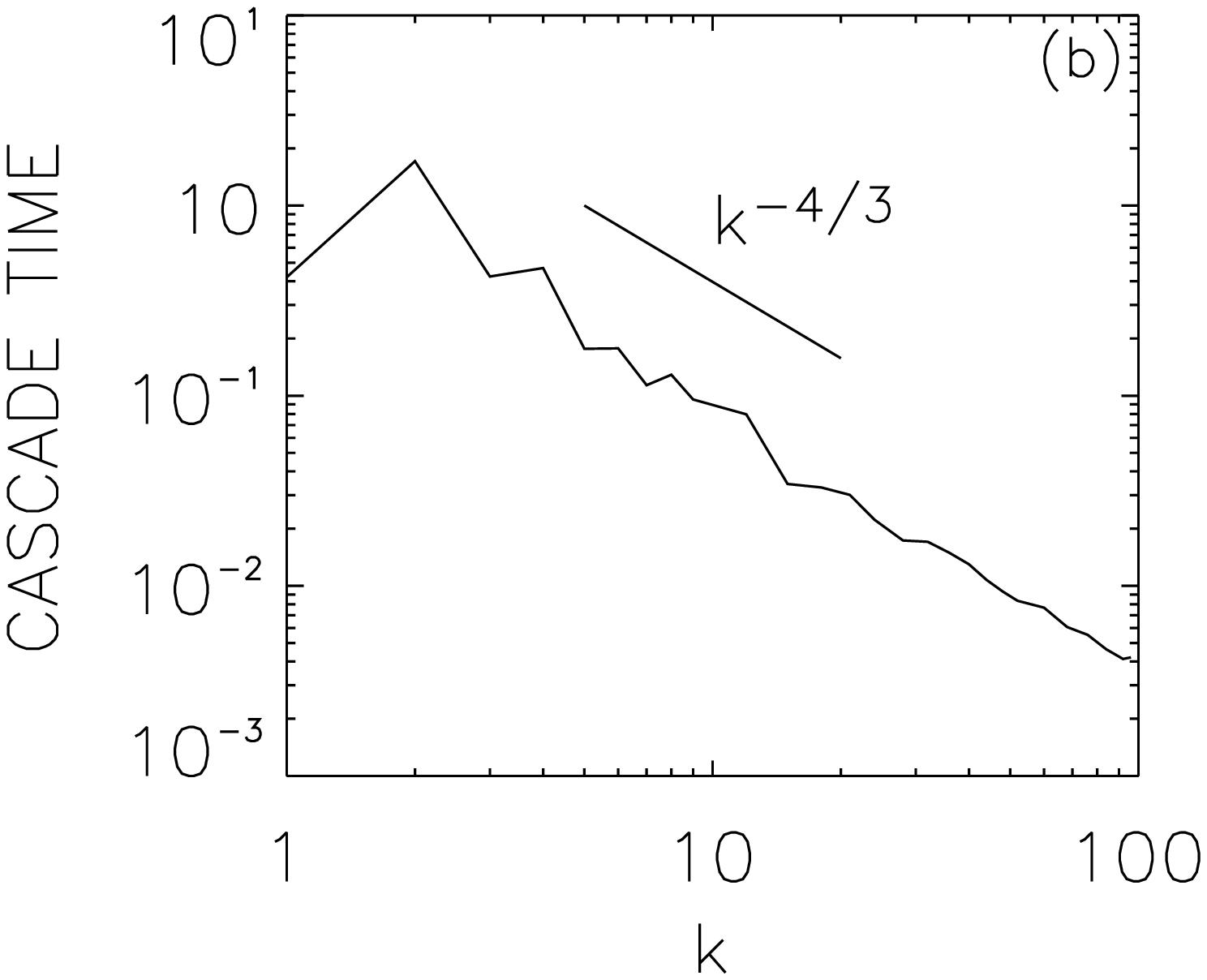}
\includegraphics[width=.30\textwidth]{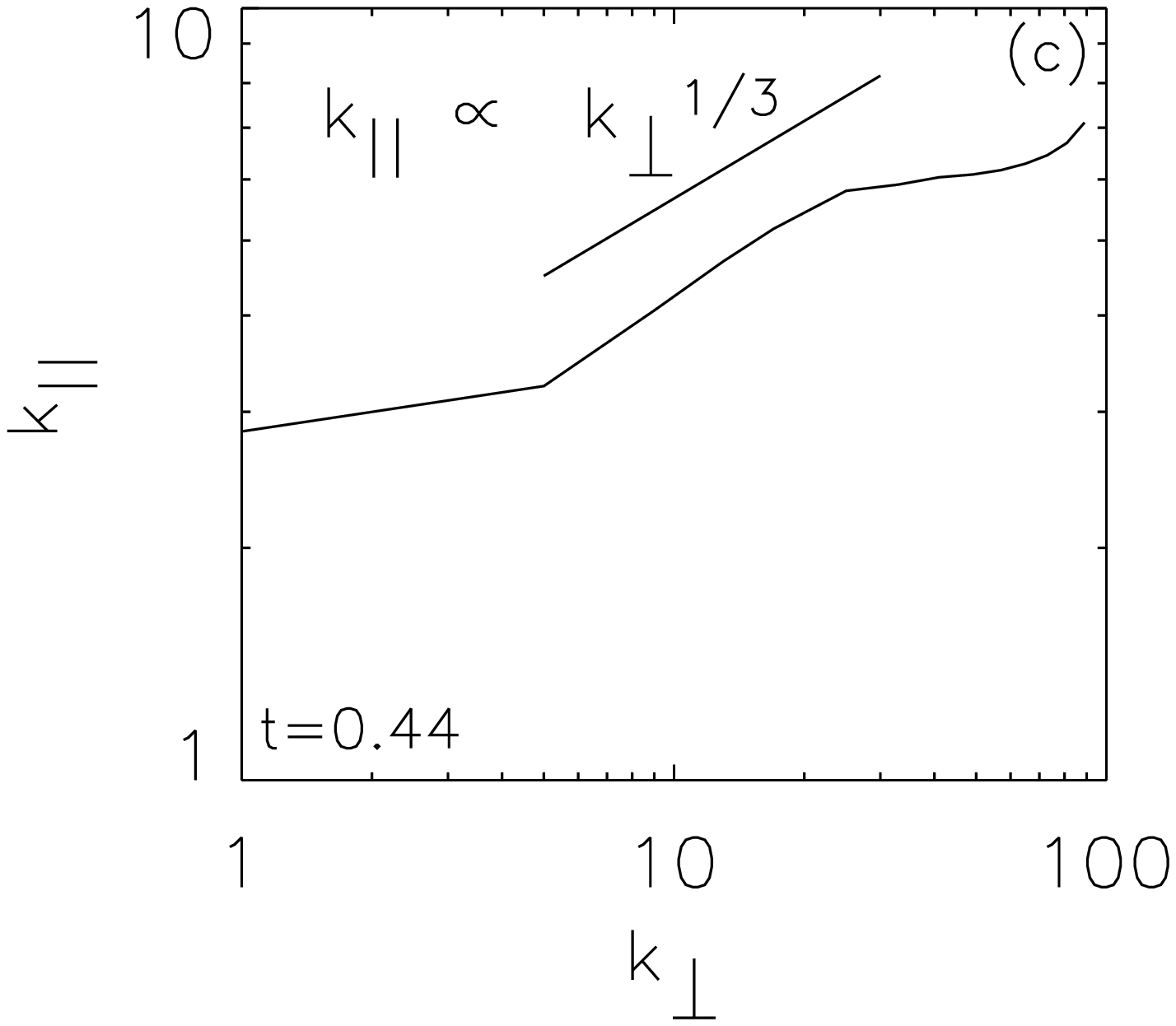}
\caption{ Energy spectra, timescale and anisotropy.
       (a)  Energy spectra for t=0 (dotted), 
       0.33 (dashed), 
             and 0.54 (solid). The spectra after $t>0.33$ are compatible
             with $E(k)\propto k^{-7/3}$.
       (b) Cascade time scales with $t_{cas} \propto k^{-4/3}$.
       (c) Anisotropy calculated from equation (\ref{kparLl}).
       }
\label{fig_sp}
\end{figure*}  


Dastgeer et al.~(2000), Ng et al.~(2003), and 
Dastgeer \& Zank (2003) numerically
studied anisotropy of 2D EMHD turbulence.
No 3D study has been reported yet.
In this paper, we study anisotropy of 3D EMHD turbulence.

{\bf Method.}--- 
We have calculated the time evolution of decaying 3D 
incompressible electron 
magnetohydrodynamic turbulence.
We have adopted a pseudospectral code to solve the normalized
incompressible EMHD equation in a periodic box of size $2\pi$:
\begin{equation}
\frac{\partial {\bf B}}{\partial t}=-
     \nabla \times \left[
    (\nabla \times {\bf B}) \times{\bf B} \right] 
    + \eta^{\prime} \nabla^{2} {\bf B},
     \label{beq}
\end{equation}
where magnetic field, time, and length are normalized by
a mean field $B_0$, the whistler time $t_w=L^2(\omega_{pe}/c)^2/\Omega_e$
($\Omega_e$= electron gyro frequency), and
a characteristic length scale $L$ 
(see, for example, Galtier \& Bhattacharjee 2003).
The resistivity $\eta^{\prime}$ in equation (\ref{beq}) is dimensionless.
The dispersion relation of a whistler waves in this normalized units
is $\omega=kk_{\|}B_0$, where $k_{\|}$ is the wave number parallel
to the (local) mean magnetic field.
The magnetic field consists of the uniform background field and a
fluctuating field: ${\bf B}= {\bf B}_0 + {\bf b}$.
The strength of
the uniform background field, $B_0$, is set to 1.
We use $288^3$ collocation points.
At $t=0$, the random magnetic field is restricted to the range
$2\leq k \leq 4$ in wavevector ({\bf k}) space.
The amplitudes of the random magnetic field at $t=0$ is $\sim 1$.
Hyperdiffusivity is
used for the diffusion terms.
The power of hyperdiffusivity
is set to 3, so that the dissipation term in the above equation
is replaced with
  $\eta_3 (\nabla^2)^3 {\bf B}$,
where $\eta_3$ is approximately $2 \times 10^{-10}$.

\section{Results}

{\bf Spectrum.}---Figure~\ref{fig_sp} shows energy spectra. 
At $t=0$ only large scale 
(i.e. small $k$)
Fourier modes are excited. The dotted curve in Figure~\ref{fig_sp} shows the
initial spectrum.
As the turbulence decays, the initial energy cascades down to small scales 
and, as a result, small scale (i.e. large $k$) modes are excited.
When the energy reaches the dissipation scale at $k>70$,
the energy spectrum decreases without changing its slope (the dashed and the 
solid curves).
The slope at this stage is around $2.20$, which is
very close to the predicted spectrum:
\begin{equation}
  E(k) \propto k^{-7/3}.
\end{equation}
However, we also note that our result is compatible with
a $k^{-2}$ spectrum.

\vspace{0.3cm}
{\bf Timescale.}---The energy spectra confirm that the 
Kolmogorov-type energy cascade model with $t_{cas}\sim l/v_l$
works fine
for EMHD turbulence.
Here we show that timescale of motion also supports the energy cascade model.

Symbolically, we can rewrite the EMHD equation in wave-vector space as follows:
    $\dot{\bf b}_{\bf k} = {\bf N}^b_{\bf k}$,
where ${\bf N}^b_{\bf k}$ represents
the nonlinear term.
We have ignored the dissipation term.
Naively speaking, we might obtain the time scale
by dividing $|{\bf b}_{\bf k}|$ by $|{\bf N}^b_{\bf k}|$.
However, to remove the effect of large scale translational motions, we 
restrict the
evaluation of the nonlinear terms to contributions coming from the
interactions between the mode at $k$ and
other modes within the range of $k/2$ and $2k$ 
(see Cho, Lazarian, \& Vishniac (2002a) for details). 
We show the result in Figure 1(b).
Our result seems to follow the timescale
\begin{equation}
  t_{cas} \sim l/v_l \sim l^{4/3} \sim k^{-4/3}.    \label{eq_tcas}
\end{equation}

\vspace{0.3cm}
{\bf Anisotropy.}---
Since ${\bf B}\cdot \nabla {\bf B} \sim B_0 <k_{\|}> b$, 
we can obtain the average parallel
wavenumber $<k_{\|}>$ from
$<k_{\|}> \sim ( |{\bf B}\cdot \nabla {\bf B}|^2/B_0^2/b^2)^{1/2}$, where
$b$ is the strength of the random magnetic field.
The quantity $1/<k_{\|}>$ is roughly the average 
parallel size of eddies at the energy
injection scale.
We can easily extend this method to smaller scale eddies.
Consider eddies of perpendicular size $l$.
Since it is reasonable to assume that the eddies are stretched along the
{\it local} mean magnetic field, the quantity 
${\bf B}_L\cdot \nabla {\bf b}_l$, not ${\bf B}_0\cdot \nabla {\bf b}_l$,
is proportional to $B_L k_{\|} b_l$, where ${\bf B}_L$ is the local mean field,
${\bf b}_l$ the fluctuating field at scale $l$, and
$k_{\|}$ the wave number 
parallel to the local mean magnetic field
(Cho, Lazarian, \& Vishniac 2002b).
Therefore, we have 
$k_{\|}(l) \sim 
( |{\bf B}_L\cdot \nabla {\bf b}_l|^2/B_L^2/b_l^2)^{1/2}$.
It is convenient to do the calculation in Fourier space:
\begin{equation}
k_{\|}(k_{\perp}) \approx \left(
     \frac{\sum_{k\leq |{\bf k}^{\prime}| <k+1}
   |\widehat{ {\bf B}_L\cdot \nabla {\bf b}_l   }|_{{\bf k}^{\prime}}^2  }
  { B_L^2 \sum_{k\leq |{\bf k}^{\prime}| <k+1}
                     |\hat{\bf b}|^2_{{\bf k}^{\prime}} }
               \right)^{1/2},    \label{kparLl}
\end{equation}
where $k_{\perp}$ 
($\sim k \sim 1/l$ when anisotropy is present) is the wave number 
perpendicular to the local mean magnetic field and 
${\bf B}_L$ is the local mean field obtained by
eliminating
modes whose perpendicular wavenumber is greater than $k/2$ ($\sim 1/(2l)$). 
The fluctuating field ${\bf b}_l$ is obtained by eliminating
modes  whose perpendicular wavenumber is less than $k/2$.
We plot the result in Figure 1(c).
The result does not show a well-defined power law between
$k_{\|}$ and $k_{\perp}$. 
However, the result is roughly compatible with
\begin{equation}
  k_{\|} \sim k_{\perp}^{1/3}.  \label{eq_ani_2}
\end{equation}
We believe this is the true anisotropy of EMHD turbulence.

We may use the second order structure function to
illustrate scale-dependent anisotropy of eddy structures 
(see Cho \& Vishniac 2000).
We calculate the second order structure
function in a local frame, which is aligned with local mean magnetic field
${\bf B_L}$:
\begin{equation}
    \mbox{SF}_2(r_{\|},r_{\perp})=<|{\bf B}({\bf x}+{\bf r}) -
                 {\bf B}({\bf x})|^2>_{avg.~over~{\bf x}},
\end{equation}
where ${\bf r}=r_{\|} {\hat {\bf r}}_{\|} +r_{\perp} {\hat {\bf r}}_{\perp}$
and ${\hat {\bf r}}_{\|}$ and ${\hat {\bf r}}_{\perp}$ are unit vectors
parallel and perpendicular to the local mean field ${\bf B_L}$, respectively.
See Cho et al. (2002a) 
and Cho \& Vishniac (2000) for the 
detailed discussion of the local frame.
We show the resulting structure function in Figure 2.


\begin{figure}[t]
\includegraphics[width=.70\columnwidth]{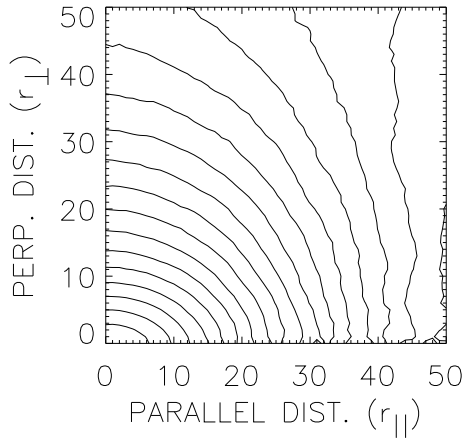}

\caption{    
          Contour plot of the second structure function 
          $SF_2(r_{\|},r_{\perp})$ at $t=0.44$, 
          where $r_{\|}$ ($r_{\perp}$) is
          distance along (perpendicular to) the $local$ mean field,
          shows existence of scale-dependent anisotropy:
          smaller eddies are more elongated.
          Although this plot is useful to illustrate scale-dependent
          anisotropy, we do not use this plot to study
          anisotropy (see text). 
       }
\label{fig_ani}
\end{figure}  


We may obtain anisotropy of eddy structures by analyzing 
the contour shapes.
However, we believe that this method will not give the true
anisotropy for EMHD.
The reason is as follows.
Roughly speaking, when energy spectrum is steeper than $k^{-3}$ 
we should not use the method.
When energy spectrum is steeper than $k^{-3}$, small scale fluctuations are
so small that the
calculation of the second-order
structure function will be dominated by the smooth change of 
the large scale field and
the second-order structure function will show $r^{2}$-scaling regardless of
the true scaling:
$<|B(x)-B(x+r)|^2> \sim <|B_L(x)-B_L(x+r)|^2> \propto  r^{2}$ when $r <  L$.
In the perpendicular direction,
$E(k_{\perp})\propto k_{\perp}^{-7/3}$ and, therefore, it is fine to use
the structure function.
However,  
we should not use the method for the {\it parallel} direction
when anisotropy is stronger than $k_{\|} \propto k_{\perp}^{2/3}$.
Suppose anisotropy scales with $k_{\|} \propto k_{\perp}^{\alpha}$.
Then the 3D energy spectrum is 
$E_{3D} (k_{\|},k_{\perp}) \propto k_{\perp}^{-10/3 -\alpha} 
  g(k_{\|}/k_{\perp}^{\alpha})$, where $g$ is a function that
describes distribution of energy along the $k_{\|}$ direction 
in Fourier space (see Goldreich \& Sridhar 1995; Cho et al. 2002a).
The 1D energy spectrum
for the {\it parallel} direction becomes
$E(k_{\|}) \propto \int k_{\perp} d\phi  dk_{\perp} 
   E_{3D} (k_{\|},k_{\perp})
   \propto k_{\|}^{-1-4/(3\alpha)}$ (Cho et al. 2002a).
Therefore, when $\alpha < 2/3$, $E(k_{\|})$ is steeper than 
$k_{\|}^{-3}$ and, thus,  we cannot use the structure function to study
anisotropy. 
In other words, even if $\alpha < 2/3$, the structure function method
will give a wrong result that $\alpha \approx 2/3$.


\begin{figure}[t]
\includegraphics[width=.80\columnwidth]{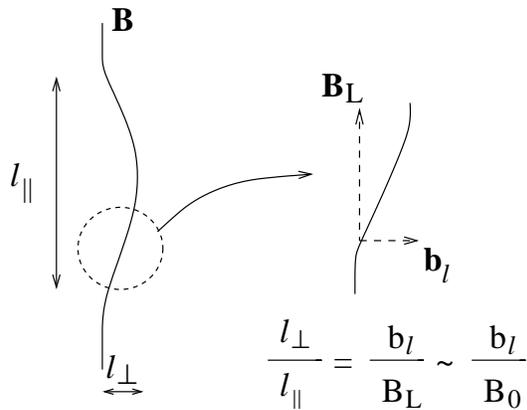}
\caption{ 
    Typical eddy shape and $b_l/B_L$ ratio.
    The degree of bending of magnetic field lines depends
    on the $b_l/B_L$ ratio.
    When we assume that $l_{\perp}$ and $l_{\|}$ represent
    typical perpendicular and parallel size of eddies at the scale $l$
    ($\sim l_{\perp}$),
    we obtain $l_{\perp}/l_{\|} \sim b_l/B_0$ and, thus, 
    $l_{\|} \sim l_{\perp}^{1/3}$, 
    we use $b_l \sim l^{2/3}$ and $B_L\sim B_0$.
}
\label{fig_geo}
\end{figure}  


\section{Discussion}

We obtained $k_{\|}\propto k_{\perp}^{1/3}$ in the previous section.
See equation (\ref{eq_ani_2}).
Here we discuss why this is reasonable.

First, the fact that the Kolmogorov-type argument works fine
for the energy spectrum implies 
the $k_{\|}\propto k_{\perp}^{1/3}$ anisotropy.
As we mentioned earlier, Biskamp et al. (1996) assumed that
the cascade time ($t_{cas}$) is the eddy turnover time 
($t_{eddy}\sim l/v_l\sim l^2/b_l$) and
obtained $E(k)\propto k^{-7/3}$, which agrees well with
numerical simulations.
See also our calculation of timescale in Figure 1(b).
We can show that, when $t_{cas}\sim t_{eddy}$, there is balance between
the eddy turnover time $t_{eddy}$ and the the whistler time 
$t_w$ (=the linear wave period=$\sim 1/(kk_{\|}B_0)$).
Suppose that we have a wave packet whose parallel size is 
$l_{\|}\sim k_{\|}^{-1}$ and
perpendicular size $l_{\perp}\sim k_{\perp}^{-1}$ ($\sim l\sim k^{-1}$ when
anisotropy is present).
This wave packet travels along the magnetic field line at the speed of
$kB_0$.
When this wave packet collides with opposite-traveling wave packets (of
similar size), the change of energy per collision is
$\Delta E  \sim ({dE}/{dt}) \Delta t  \sim 
(k^2 b_l^3)({l_{\|}}/({kB_0}))$, where we used equation (\ref{beq}) to estimate
${dE}/{dt}$.
Therefore, 
\begin{equation}
   \Delta E/E \sim (k b_l)/(k_{\|}B_0) \sim t_w/t_{eddy}.  \label{eq_dee}
\end{equation}
If $\Delta E /E < 1$, then many collisions are required
to make $\Delta E /E \sim 1$.
These collisions are incoherent. Then
$(t_{eddy}/t_w)^2$ collisions are necessary to complete the cascade.
This means that
the energy cascade time by collision is 
\begin{equation}
 t_{cas,coll} \sim  (t_{eddy}/t_w)^2 t_w.
\end{equation}
If  $t_{cas,coll} \sim t_{eddy}$,
this equation implies $t_w \sim t_{eddy}$.
Therefore, when $t_{cas} \sim t_{eddy}$, the turbulence is strong 
(i.e. $\Delta E/E \sim 1$)  
\footnote{
   The reverse is also true: if turbulence is strong,
   then $\Delta E/E \sim 1$ by definition and, therefore,
   energy cascade occurs during one wave
   period, which is equal to $t_{eddy}$ (see equation (\ref{eq_dee})).
   Thus, one can show that
   three expressions, {\it Kolmogorov-type scaling for cascade time}
   ($t_{cas}\sim l/v_l$), {\it critical balance} ($t_{eddy}\sim t_w$), and
   {\it strong turbulence} ($\Delta E/E \sim 1$), are all equivalent
   (C. Thompson, private communication),
   as in {\it ordinary} MHD turbulence.
   See Goldreich \& Sridhar (1995) 
   for the original discussion of the critical balance and anisotropy of
   {\it ordinary} MHD turbulence.
}
and there is a balance between 
$t_{eddy}\sim 1/(k^2 b_l)$ and $t_w\sim 1/(kk_{\|}B_0)$.
Our results clearly show that the critical balance factor 
$\xi \equiv t_{w}/t_{eddy}$
is constant (note that $b_l \sim l^{2/3}$ and $k_{\|}\propto k^{1/3}$).
Using $b_l\sim k^{-2/3}$, we obtain $k_{\|}\propto k_{\perp}^{1/3}$.
The uncertainty relation is another way to express this balance.
The usual uncertainty relation between time and energy ($\sim$frequency) is 
$\Delta t \propto 1/\Delta \omega$.
Since $\Delta t \sim t_{cas} \sim k^{-4/3}$ 
and $\Delta \omega \sim B_0kk_{\|}$,
we obtain the same anisotropy scaling.

Second, 
a simple geometric argument in Fig.~\ref{fig_geo} suggests
the $k_{\|}\propto k_{\perp}^{1/3}$ anisotropy.
Suppose that Fig.~\ref{fig_geo} shows
a typical shape of magnetic field lines at scale $l$.
Then, we can assume that $l_{\perp}/l_{\|} \sim b_l/B_L$, where
$b_l$ is the fluctuating field at the scale $l$ and $B_L$ is
the local mean field ($B_L \sim B_0$).
If we assume $l_{\perp}$ and $l_{\|}$ represent
typical shape of eddies at the scale $l$ ($\sim l_{\perp}$), we
have
\begin{equation}
  k_{\|}/k_{\perp} \sim b_l/B_0.  \label{eq:geo}
\end{equation}
In this interpretation, smaller eddies are more elongated because
they have a smaller $b_l/B_0$ ratio.
Note that the relation in
equation (\ref{eq:geo}) 
is in fact equivalent to the strong turbulence condition 
($\Delta E/E \sim 1$; see equation (\ref{eq_dee})),
which means energy cascade occurs over one wave period. 
{}From  equation (\ref{eq_dee}), 
we can show that
$t_{cas}\sim l/v_l \sim l^2/b_l$.
Combining this with the constancy of spectral energy cascade rate 
($b_l^2/t_{cas}$=constant),
we obtain $b_l \propto k^{-2/3}$ 
and, thus, the $k_{\|}\propto k_{\perp}^{1/3}$ anisotropy.

One peculiar feature of EMHD cascade is that there is
balance between parallel and perpendicular cascade by
self-interaction.
The resonance relations of whistler turbulence,
\begin{equation}
  {\bf k}_1 + {\bf k}_2 = {\bf k}_3 \mbox{~~and~~}
  k_1 |k_{\|,1}| +k_2 |k_{\|,2}| = k_3 |k_{\|,3}|,
  \label{eq:reso1}
\end{equation}
allow self-interaction.
We can show that the parallel cascade by self interaction is as fast as 
the perpendicular cascade when
 $k_{\|}\propto k_{\perp}^{1/3}$. 
Ng et al. (2003) performed a numerical experiment
and
found that cascade by self-interaction is as fast as
that by collision with opposite-traveling wave packets.
Combined together,
these facts imply that anisotropy cannot be
stronger than  $k_{\|}\propto k_{\perp}^{1/3}$ in EMHD.
We will discuss this in detail in a forthcoming paper.

We have focused on anisotropy of 3D EMHD turbulence.
However, it is fair to mention earlier 2D results.
Dastgeer et al. (2000) numerically studied
2D EMHD turbulence and found numerical evidence that
the turbulence is anisotropic in the presence of a mean field.
Dastgeer \& Zank (2003) showed that the {\it global}
anisotropy of    
decaying EMHD turbulence scales linearly with the inverse of
mean magnetic field strength, $1/B_0$.
Ng et al. (2003)
found local anisotropy of $k_{\|} \propto k_{\perp}^{3/4}$
from a 2D EMHD simulation. 
We will discuss the 2D versus 3D case  elsewhere.


Conventional
MHD turbulence can be
decomposed into cascades of Alfven, fast and slow modes 
(Goldreich \& Sridhar 1995; Cho \& Lazarian 2002). 
While
fast and slow modes get damped at larger scales, Alfvenic modes
can cascade down to the proton Larmor radius scale. 
These Alfven modes excite whistler modes below the scale.
Quataert \& Gruzinov (1999) discussed transition from
conventional Alfvenic MHD cascade to EMHD cascade in ADAFs
and argued that
anisotropy of EMHD turbulence is important for electron
heating in ADAFs.
They showed that, depending on the anisotropy of the resulting
whistler turbulence, the whistler cascade will or will not transfer 
most of its energy to electrons.
Whether or not whistler turbulence is anisotropic
is an important question for the physics of accretion disks 
and gamma-ray bursts.
The strong anisotropy of EMHD turbulence makes 
heating of protons by EMHD turbulence rather
difficult in astrophysical plasmas with $\beta \sim 1$, where
$\beta$ is the gas to magnetic pressure ratio (see Quataert \& Gruzinov 1999).

\section{Summary}

Using a numerical simulation, we have studied 3D EMHD turbulence.
 We have found that
energy 
cascade time scales with $k^{-4/3}$.
We have calculated anisotropy and found
$ k_{\|}\propto k_{\perp}^{1/3}$.
We have discussed why this is reasonable.

\begin{acknowledgments}  
We thank Chris Thompson for many valuable comments and 
constructive suggestions. 
J.C. thanks Elliot Quataert for emphasizing the importance of the
critical balance. 
This work was partially supported by NCSA
under AST030023.
A.L acknowledges NSF grant AST 0307869 ATM 0312282 and the NSF 
Center
for Magnetic Self Organization in Laboratory and Astrophysical Plasmas.
\end{acknowledgments}


\begin{thebibliography}{8.}
\addcontentsline{toc}{section}{References}


\bibitem{AviBE98}  Avinash, K., Bulanov, S. V., Esirkepov, T., Kaw, P., 
          Pegoraro, F., Sasorov, P. S., \& Sen, A.  
          1998, Phys. Plasmas 5, 2849

\bibitem{BisSD95} Biskamp, D., Schwarz, E.,  \& Drake, J. F.
          1995, Phys. Rev. Lett. 75, 3850
\bibitem{BisSD96} Biskamp, D., Schwarz, E.,  \& Drake, J. F.
          1996, Phys. Rev. Lett., 76, 1264
\bibitem{BisSZ99} Biskamp, D., Schwarz, E., Zeiler, A., Celani, A.,  \&
                  Drake, J. F. 1999, Phys. Plasmas, 6, 751
\bibitem{BulPS92} Bulanov, S. V., Pegoraro, F., \& Sakharov, A. S.
           1992, Phys. Fluids B, 4, 2499




\bibitem{ChoLV02a} Cho, J., Lazarian, A., \& Vishniac, E. 2002a, ApJ,
                  564, 291 
\bibitem{ChoLV02b} Cho, J., Lazarian, A., \& Vishniac, E. 2002b, ApJ,
                  566, L49

\bibitem{ChoL02} Cho, J. \& Lazarian, A. 2002, Phys. Rev. Lett., 88, 245001
\bibitem{ChoV00} Cho, J. \& Vishniac, E. 2000, ApJ, 539, 273 

\bibitem{CumAZ}  Cumming, A., Arras, P., \& Zweibel, E. 2004, ApJ, submitted
                 (astro-ph/0402392)



\bibitem{DasDK00} Dastgeer, S.,  Das, A., Kaw, P., \& Diamond, P. 2000,
                  Phys. Plasmas, 7, 571
\bibitem{DasZ03} Dastgeer, S. \& Zank, G. P. 2003, ApJ, 599, 715



\bibitem{GalB03} Galtier, S. \& Bhattacharjee, A. 2003, Phys. Plasmas, 
                 10, 3065

\bibitem{GolR92} Goldreich, P. \& Reisenegger, A. 1992, ApJ, 395, 250
         
\bibitem{GolS95} Goldreich, P. \& Sridhar, S. 1995, ApJ, 438, 763


\bibitem{KinCY90} Kingsep, A. S., Chukbar, K. V., \& Yan'kov, V. V. 1990, in 
   {\it Reviews of Plasma Physics}, Vol. 16 (Consultants Bureau, New York)


\bibitem{NgBG03} Ng, C. S., Bhattacharjee, A., Germaschewski, K., \&
                 Galtier, S. 2003, Phys. Plasmas, 10, 1954


\bibitem{QuaG99} Quataert, E. \& Gruzinov, A. 1999, ApJ, 520, 248









\end{thebibliography}
\end{document}